\documentclass[10pt,conference]{IEEEtran}
\IEEEoverridecommandlockouts
\usepackage{cite}
\usepackage{amsmath,amssymb,amsfonts}
\usepackage{graphicx}
\usepackage{textcomp}
\usepackage{xcolor}

\usepackage{amsmath, amssymb, amsthm,algorithm}
\usepackage[latin1]{inputenc}
\usepackage{xcolor}
\usepackage{latexsym}
\usepackage{graphics,epsfig}
\usepackage{amsfonts}
\usepackage{booktabs}
\usepackage{color}
\usepackage{multirow}
\usepackage[justification=centering]{caption}
\usepackage{graphicx}
\usepackage{adjustbox}
\usepackage{kantlipsum}
\usepackage[caption=false,font=footnotesize]{subfig}
\usepackage{cases}
\usepackage{url}
\usepackage{tabu}
\usepackage{makecell}
\usepackage{soul}

\usepackage{array}
\usepackage{algpseudocode}
\usepackage[justification=centering]{caption}

\usepackage[caption=false,font=footnotesize]{subfig}
\usepackage[T1]{fontenc}
\usepackage{mwe}
\usepackage{subfig}
\usepackage{mathtools,lipsum,cuted}

\usepackage[english]{babel}
\usepackage{verbatim}

\usepackage[english]{babel}
\algrenewcommand{\algorithmiccomment}[1]{\hspace{-20px}$\rightarrow$ #1}

\def\BibTeX{{\rm B\kern-.05em{\sc i\kern-.025em b}\kern-.08em
    T\kern-.1667em\lower.7ex\hbox{E}\kern-.125emX}}
\begin{document}

\title{Parallel Downlink Data Distribution \\in Indoor Multi-hop THz Networks}
\author{\IEEEauthorblockN{Cao Vien Phung, Andre Drummond
 and Admela Jukan}
\IEEEauthorblockA{Technische Universit\"at Braunschweig, Germany\\
Email: \{c.phung, andre.drummond, a.jukan\}@tu-bs.de}}
\maketitle

\begin{abstract}
The emerging dynamic Virtual Reality (VR) applications are the best candidate applications in high bandwidth indoor Terahertz (THz) wireless networks, with the Reconfigurable Intelligent Surface (RIS) devices presenting a breakthrough solution in extending the typically short THz communication range and alleviating line-of-sight link blockages.  In future smart factories, it is envisioned that factory workers will use VR devices via VR application data with high quality resolution, while transmitting over THz links and RIS devices, enabled by the Mobile Edge Computing (MEC) capabilities. Since indoor RIS placement is static, whereas VR users move and send multiple VR data download requests simultaneously, there is a challenge of proper network load balancing, which if unaddressed can result in poor resource utilization and low throughput. To address this challenge, we propose a parallel  downlink data distribution system and develop multi-criteria optimization solutions that can improve throughput, while transmitting each downlink data flow over a set of possible paths between source and destination devices. The results show that the proposed system can enhance the performance in terms of throughput benefit, as compared to the system using one serial download link distribution.
 \end{abstract}
\begin{IEEEkeywords}
Smart factory; RIS-assisted THz networks; linear programming.
\end{IEEEkeywords}
\vspace{-0.3 cm}
\section{Introduction} \label{intro}
Communications in Terahertz (THz) band are today considered a promising technology to enable indoor high-speed networking. To alleviate the issues of short reach, and physical layer impairments, human blockage \cite{9565222}, etc., the technology called Reconfigurable Intelligent Surfaces (RISs) are proposed as a useful solution for multi-hop communications \cite{9686640,8466374,https://doi.org/10.48550/arxiv.2103.15154}. Many indoor wireless network scenarios can make use of THz communications combined with RIS. One such typical application is the so-called smart factory, e.g., on how to assemble a particular product or a component and hands-free operations through the real-time placement of images onto the real working environment \cite{boeing2018} \cite{https://doi.org/10.48550/arxiv.2205.13057}.  

\begin{figure}[ht]
 \vspace{0.1 cm}
  \centering
  {\includegraphics[ width=8.5cm, height=6.7cm]{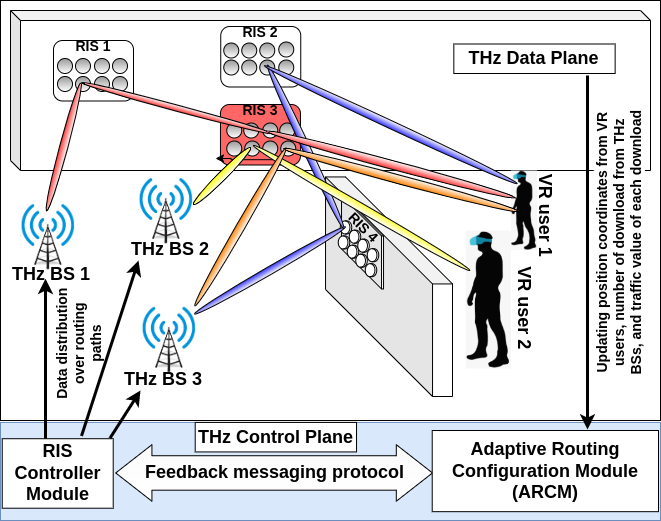}}
  \caption{A case study of a RIS-assisted multi-hop THz network in a smart factory.} \label{arch}
  \vspace{-0.68 cm}
  \end{figure}

Let us consider a case study of smart factories shown in Fig. \ref{arch}. Here,  workers use Augmented or Virtual Reality (AR, VR) and high resolution receivers for data from multiple static Base Stations (BSs) (as transmitters). The devices are connected to BSs that provide the VR application data with high quality resolution, which requires high data rates \cite{VRapps}. Multiple static RISs are equipped with MEC capabilities with the transmission queue shared by all VR users \cite{9149411}. As RIS $3$ serves many data transmission services at the same time, it can become overloaded. The scheduling of each RIS to serve many VR users can overload the MEC server,  resulting in high delay, or even service blockage in case of the MEC server with a limited queuing length assigned for each RIS. This occurs when the dynamic VR users reach a RIS in the nearest transmission distance, although other RISs are also free, which is inefficient and lowers the overall system throughput. Since indoor RIS placement is static, whereas factory workers move and send multiple VR data download requests simultaneously, the challenge is to balance the RIS utilization.

In this paper,  we address the problem of optimal load balancing in multi-hop THz and RIS-assisted networks motivated by the said smart factory case study. We conceptualize the THz/RIS network system using the  parallel distribution of downlink data in indoor networks. We distinguish architecturally between two planes: data and routing control planes. The data plane is concerned with transmitting, distributing and receiving the data, while the control plane controls routing configurations to distribute the data on the computed path. We use a linear optimization program (LP) to optimize RIS allocations, under the constraints of the queue length of the RISs equipped with MEC capabilities. Our solution can maximize the throughput by using the so-called Parallel Downlink Distribution Technique (PDDT) of each  VR application over different paths, thus offering a better system performance overall. The results show that the proposed system can improve the throughput, as compared with the system using one Serial Downlink Distribution Technique (SDDT) for the traffic chunks of each VR application data download.

The remainder of this paper is organized as follows. Section \ref{RW} presents the related work. Section \ref{sysde} presents the THz system with parallel data distribution. In section \ref{Numerre}, we show numerical results. Finally, section \ref{concl} concludes the paper.

\section{Related work}\label{RW}

Study \cite{9410457} proposes  multi-hop RIS-assisted THz networks, which can be applied for our smart factory case study, when THz BSs transfer their data to VR users in the far areas with obstacles causing link blockages. Several authors discussed that as RISs can be configured for the certain phase shifts, the reflecting signals can be constructively focused, steered, and enhanced the received signal power towards the desired user, or destructively at non-intended users with the aim of suppressing the co-channel interference in their vicinity \cite{9136592,8466374,8910627,8981888}. Considering these studies, we assume a RIS-assisted THz network that can avoid co-channel interference.

In \cite{9149411}, the authors discussed that each RIS is equipped with MEC capabilities with the transmission queue shared to all VR users, and all VR users are scheduled and processed sequentially VR applications by that RIS with using the time division multiplexing. This operation of RIS is applied to our scenario. However, serving so many VR users can result in poor resource utilization and low throughput. We address this challenge by using the proposed PDDT method, which can balance load among RISs. Several related ideas have been already proposed which we use here. For example, the authors in \cite{8845312} used multiple propagation paths to build secure THz communications. We use the throughput definition from here \cite{5438834}; throughput is defined as a maximum multiplier $\lambda$ such that all demands with the traffic values multiplied by $\lambda$ can be feasibly distributed in the network. In this way, the paper studied how to maximize the throughput, i.e., maximize $\lambda$, by finding the paths according to a certain criteria, in their case by finding paths that are better suitable for network coding. We adapt that value $\lambda$ to our problem of finding the paths with the objective of optimal load balancing among RISs for all VR downlink distributions of data in RIS-assisted THz networks.

Currently, many RIS-assisted THz network problems can be formulated as system performance optimization problems, such as sum-rate maximization problem of user equipments \cite{9672716}, maximizing the long-term total quality of experience under the downlink transmission latency constraint in continuous time slots \cite{9565222}, etc. Ideally, the optimization problem can be solved by using a LP. In this paper, we seek to maximize the RIS-assisted THz network throughput by distributing the VR data for each download over parallel paths. The RIS and transmitter (e.g., BS) are connected to a RIS controller to exchange information \cite{8910627,9209937}, e.g., phase information. We use the concept of the RIS controller to distribute the data from each VR application data download to transfer over a set of parallel links, involving RISs. 

The main contribution of this paper is a technique to optimize load balancing among RISs with MEC capabilities and improve throughput in multi-hop THz networks. The novelty is in the parallel downlink data distribution. The model is akin to multipath routing and propagation, it however also considers parallel download transmission from multiple transmitters, which is new perspective relevant to smart factories. 

\section{System design}\label{sysde}

 \subsection{THz data plane}\label{data_plane}
The THz data plane includes a $B$ set of THz BSs (source nodes), a $R$ set of RISs (intermediate nodes), and a $V$ set of VR receiver devices (destination nodes). In our case study, factory workers (VR users) using VR devices as receivers. Each RIS has a $N$ set of reflecting elements. We consider active RISs \cite{9377648} that can reflect and amplify the incident signals. Furthermore, we assume each RIS is equipped with MEC \cite{9149411} capabilities to schedule and process the queue of VR applications sequentially, as in \cite{9149411}. Also, we assume each RIS has a phase shift vector, with respect to another network node as in \cite{9149411}. We consider each THz BS exploiting the parallel transmission technique as in \cite{8845312}, where its channels use different center frequencies. THz BSs are only responsible for transmitting VR application data for download, while VR users are only responsible for receiving data from BSs. The intermediate nodes are only RISs with both transmitting and receiving data.  THz BSs and RISs are static, while VR users  move. Assume that with RISs, they can suppress the co-channel interference or creating  "interference-free zone" in their vicinity as discussed in \cite{9136592,8466374,8910627,8981888}. We assume each VR user manages the factory work at fixed physical places with different directions. In other words, we assume no link blockages due to the possible configurations of RISs. The selection of routing configurations and data distribution on the different paths will be discussed in subsection \ref{routing_plane} and \ref{LP}.


\subsection{THz control plane}\label{routing_plane}
In THz control plane, we propose to introduce two main modules, referred to as RIS Controller Module and Adaptive Routing Configuration Module (ARCM). In between, a out-of-band signaling feedback messaging protocol is assumed for exchanging the routing configurations and data distribution on those routing paths via a feedback channel. The RIS controller module focuses on controlling the data transferred over the system, while the ARCM is responsible for computing the routing paths suitable for parallel data distribution.  

To generate the new downlink configurations,  ARCM is required to update the current network connectivity information (i.e.,   topology). We assume that ARCM only updates new configurations when VR users move and network topology changes. There is a timing issue on how to actuate these changes. First, not all user movements result is system reconfigurations. For instance, when users move fast, the system is generally able to keep up with the most actual states keeping too frequent updates rather obsolete. Also the related signaling traffic in the control plane would be excessively large. To address this issue, we propose that the position coordinates of each user are periodically updated at a constant time interval $t$ by ARCM. Assume that an average moving speed of a worker is at $1$ m/s, i.e., it would take $0.5$ s with a footstep of $0.5$ m. Hence, to estimate the moving direction of each VR user and the next place to move, we recommend periodically updating its coordinates at the time intervals of $t=0.5$ s.

Based on the assumption that the position coordinate of each VR user is periodically updated, ARCM module can characterize the user status in the control plane, i.e., \emph{moving} or \emph{standing} at a certain position. As it was assumed and described in Section \ref{data_plane}, each VR user manages the work at some fixed places with different directions. Hence, based on its position coordinates periodically updated, ARCM can estimate the moving direction such that the next location to transmit the data.  Also, we assume that workers' movement is less frequent than the ability of the system to update, which is typically the case in factory settings where users do not move frequently but occasionally between a few specific locations. As a result, ARCM will collect the newly estimated positions of VR users, and then store all possible connectivity options among THz BSs, RISs and the related VR devices. With every new network topplogy, we can find the set of possible shortest paths for transmitting over downlinks. The specific algorithm to this end will be presented in Section \ref{LP}.

Let us illustrate the idea of PDDT with help of ARCM. Assume that a virtual node is connected to all THz BSs. In Fig. \ref{arch}, a VR application traffic $\alpha$ is assumed transferring from a $\mathbb{S}_{\alpha}$ set of multiple sources, e.g., $|\mathbb{S}_{\alpha}|=2$ $s_{\alpha}$ sources, $\mathbb{S}_{\alpha} = \{$THz BS $1$, THz BS $3\}$, to one destination node $d_{\alpha}= $ VR user $1$ in parallel. Each source node $s_{\alpha}=$ THz BS $1$ and source node $s_{\alpha}=$ THz BS $3$ are assigned the corresponding traffic chunk $t_{s_{\alpha}}$ split from the  application traffic $\alpha$. The THz BS $1$ sends its traffic chunk $t_{s_{\alpha}}$ to the VR user $1$ with a downlink of $3$ transmission hops (red link: THz BS $1$ $\rightarrow$ RIS $1$ $\rightarrow$ RIS $3$ $\rightarrow$ VR user $1$). Hence, we have $|\mathbb{P}_{s_{\alpha}}|=1$ shortest path between THz BS $1$ and VR user $1$. Let us assume that  RIS $3$ is nearly overloaded in relaying various links.  THz BS $3$ needs to transmit its traffic chunk $t_{s_{\alpha}}$ to VR user $1$.  THz BS $3$ may not be able to use RIS $3$ (orange link) in its downlink selection in case of overloaded RIS $3$, when THz BS $3$ tries to send its data to VR user $1$ with a downlink of $2$ transmission hops (THz BS $3$ $\rightarrow$ RIS $3$ $\rightarrow$  VR user $1$). In this case, we use the PDDT. In addition to one downlink, THz BS $3$ establishes another downlink but without RIS $3$: THz BS $3$ $\rightarrow$ RIS $4$ $\rightarrow$ RIS $2$ $\rightarrow$ VR user $1$ (blue links with $3$ transmission hops). Hence, we have $|\mathbb{P}_{s_{\alpha}}|=2$ possible shortest paths between THz BS $3$ and VR user $1$. The THz BS $3$ will  distribute in parallel its traffic chunk $t_{s_{\alpha}}$ over those paths so that the application traffic $\alpha$ can be sent to VR user $1$. We observe these downlinks are not necessarily the shortest ones but allow avoiding overloaded RISs, thus offering a better throughput. 

The parallel downlink data distribution is decided by ARCM, and then sent out to RIS controller for configurations. After the RIS controller receives new configurations, it will signal THz BSs and RISs to transmitting VR application download split. Splitting the data at the transmitter side is not trivial and neither is assembling of the parallelized data at the receiver.  It requires consideration of the differential delay which is out of scope here, since we limit our scope to RIS load balancing and assume that all downlink options are equally valid and do not cause differential delay due to extremely short indoor distances with comparable delays. 

\subsection{Problem formulation}\label{LP}
 \begin{table}[t!]
  \centering
    \vspace{0.2 cm}
  \caption{List of input parameter and variable notations.}
  \label{tab:table2}
  \begin{tabular}{ll}
    \toprule
 Input param. & Meaning\\
    \midrule
    $E_{out}(br)$ & Set of outgoing links of THz BS node $b \in B$, \\
   & or RIS node $r \in R$, i.e., $br \in \{ B \cup R \}$.\\
   $E_{in}(rv)$ & Set of incoming links of RIS node $r \in R$,\\
   & or VR user $v \in V$, i.e., $rv \in \{ R \cup V \}$.\\
   $S_{r}$ & Average data transmission speed of RIS node $r \in R$.\\
   $A$ & VR application traffic set: Each application traffic \\
   & $\alpha \in A$ is transferred from a $\mathbb{S}_{\alpha}$ set of multiple  \\
   & $s_{\alpha}$ sources to one destination node $d_{\alpha}$ in parallel.\\
  $t_{s_{\alpha}}$ & Each source node $s_{\alpha}$ is assigned the traffic chunk $t_{s_{\alpha}}$\\
  & split from the VR application traffic $\alpha \in A$. \\
   $\mathbb{P}_{s_{\alpha}}$ & Each traffic chunk $t_{s_{\alpha}}$ can be parallelly distributed \\
   & over the $\mathbb{P}_{s_{\alpha}}$ set of possible shortest paths between\\
   &  the source node $s_{\alpha}$ and destination node $d_{\alpha}$.\\
   $L$ & Queuing length of each RIS node equipped with MEC.\\
    \toprule
  Variables & Meaning\\
  \midrule
 $y_r$ & Traffic variable at RIS node $r\in R$.\\
 $\lambda$ & Throughput Controlling Coefficient (TCC).\\
 $f_{s_{\alpha}}^P$ & Traffic variable sent from source node $s_{\alpha}$ \\
& over path $P \in \mathbb{P}_{s_{\alpha}}$.\\
     \bottomrule
  \end{tabular}\vspace{-0.5cm}
\end{table}

We now present the optimization method based on linear programming that can be used to distribute traffic over parallel downlinks in RIS-assisted THz networks. We give a general optimization model with the overall traffic sent from multiple THz BSs to multiple VR users. All input parameters for the model are summarized in Table \ref{tab:table2}.

For simplicity, we assume that the reflecting signals of each RIS $r$ which reach to the same destination (e.g., another RIS $r^{,}$ or VR user $v$) are considered as one outgoing wireless link $(r,r^{,}v)$ for that RIS $r$, whereby $C_{(r,r^{,}v)}(Z)$  presented in Eq. \eqref{capacitychanRIS} is the total data transmission speed from the RIS $r$ to another RIS $r^{,}$ or VR user $v$ \cite{9149411}, and $Z$ is the set of reflecting elements of RIS $r$ used for reflecting the signals from the RIS $r$ to the related network node $r^{,}v$ over the link $(r,r^{,}v)$. Assume that the current network topology is updated by ARCM, based on the updates of position coordinates from VR users. This network structure is used to choose the $\mathbb{P}_{s_{\alpha}}$ set of possible shortest paths from the source node $s_{\alpha}$ and destination node $d_{\alpha}$ for each traffic chunk $t_{s_{\alpha}}$. The ARCM will also update the VR application traffic set $A$ from THz BSs and the $t_{s_{\alpha}}$ traffic chunks assigned to the  $s_{\alpha}$ source nodes. 

For a given routing scheme, and considering the transmission impairments of each RIS involved (Eq. \eqref{capacitychanRIS}), we define TCC as a maximum multiplier $\lambda$ such that all $t_{s_{\alpha}}$ traffic chunks (assigned to the $s_{\alpha}$ sources) multiplied by $\lambda$ can transmitted without blocking. Then, the downlink allocation in THz networks with minimum RIS-overload to maximize throughput can be formulated based on an LP model, which we adapt from \cite{5438834}. The outputs of LP represent the best downlink choice for each traffic chunk $t_{s_{\alpha}}$ from the $f_{s_{\alpha}}^P$ traffic variables  with the variable $\lambda$ so that each traffic chunk $t_{s_{\alpha}}$ multiplied by $\lambda$ can be feasibly transmitted in the network. The objective function is defined as:
\begin{equation} \nonumber
\text{Maximize} \; \lambda 
\end{equation}
\begin{equation}
\begin{split}
\text{subject to} \sum\limits_{P \in \mathbb{P}_{s_{\alpha}}}f_{s_{\alpha}}^P = t_{s_{\alpha}} \cdot \lambda, \; \forall \; \alpha \in A, \; s_{\alpha} \in \mathbb{S}_{\alpha}. \label{cont1}
\end{split}
\end{equation}
\begin{equation}\label{cont4}
y_r = \sum_{e \in E_{in}(r)} \; \sum_{\alpha \in A} \; \sum_{s_{\alpha} \in \mathbb{S}_{\alpha}} \; \sum_{P \in \mathbb{P}_{s_{\alpha}}, e \in P} f_{s_{\alpha}}^P, \forall \;  r \in R.
\end{equation}
\begin{equation}\label{cont7}
\begin{split}
y_r - (S_r \cdot \tau) \leq L , \; \forall r \in R.
\end{split}
\end{equation}

Constraint \eqref{cont1} defines that for each source node $s_{\alpha}$ assigned the traffic chunk $t_{s_{\alpha}}$ (split from the application traffic $\alpha \in A$), the total amount of traffic $\sum\limits_{P \in \mathbb{P}_{s_{\alpha}}}f_{s_{\alpha}}^P$, transferred from that source node $s_{\alpha}$ to the destination node $d_{\alpha}$ from the $\mathbb{P}_{s_{\alpha}}$ set of possible shortest paths, is equal to that $t_{s_{\alpha}}$ multiplied by $\lambda$.



Constraint \eqref{cont4} represents that for each RIS node $r \in R$, the total amount of traffic $y_r$ is transferred through that RIS node. This amount of transit traffic is from all traffic requests $\alpha \in A$ that are actually using incoming links $e \in E_{in}(r)$.

Constraint \eqref{cont7} states that for each RIS $r$,  the amount of remaining traffic after the time interval $\tau$ observed should be less than or equal to the queuing length $L$ of that RIS to avoid the overload. $S_r$ is the average transmission speed of each RIS $r \in R$, where each outgoing link is served at a time by the RIS \cite{9149411}. Therefore, $S_r$ is given by:
\begin{equation}\label{cont6}
\begin{split}
S_r = \frac{1}{|X|} \sum_{|X|} C_{(r,r^{,}v)}(Z), \forall r \in R.
\end{split}
\end{equation}

The mathematical notations of Eq. \eqref{cont6} are explained as follows. $S_r$ is calculated from  $|X|$ outgoing links between the RIS node $r$ and the other RIS nodes or VR users (nodes $r^{,}v \in \{ R \cup V \}$), whereby $X$ is given by:
\begin{equation}\label{calX}
\begin{split}
X  = \{ (r,r^{,}v) \in E_{out}(r) \; & | \; \exists \; (r,r^{,}v) \in P, \\
& \forall \; \alpha \in A, s_{\alpha} \in \mathbb{S}_{\alpha}, P \in \mathbb{P}_{s_{\alpha}}  \}.
\end{split}
\end{equation}
For each outgoing link $(r,r^{,}v) \in X$, we have the data transmission speed $C_{(r,r^{,}v)}(Z)$ from the RIS $r$ to another node (RIS $r^{,}$ or VR user $v$, i.e., node $r^{,}v \in \{ R \cup V \}$), given by \cite{9149411}:
\begin{equation}\label{capacitychanRIS}
\begin{split}
& C_{(r,r^{,}v)}(Z) = W \cdot log_2  \\
& \left(1 + \frac{P_{RIS} \cdot H_{(r,r^{,}v)} \cdot   \sum_{n_e \in Z} \left| e^{j(\phi_{(r,n_e,r^{,}v)} - \psi_{(r,n_e,r^{,}v)})}  \right|^2}{\frac{W \cdot c^2}{4 \cdot \pi \cdot f^2}k_B \cdot T_0+P_{RIS} \sum_{i \in I} A_i +P_{BS} \sum_{i \in J} A_i}  \right).
\end{split}
\end{equation}
Eq. \eqref{capacitychanRIS} is explained as follows: According to \cite{9149411,8763780}, $ (\frac{W \cdot c^2}{4 \cdot \pi \cdot f^2}k_B \cdot T_0+P_{RIS} \sum_{i \in I} A_i +P_{BS} \sum_{i \in J} A_i  )$ is the total noise power of the molecular absorption and the Johnson-Nyquist noise generated by thermal agitation of electrons in conductors, whereby $W$ is the bandwidth, $c=3\cdot 10^8$ m/s, $\pi=3.14$, $f$ is the operating frequency, $k_B$ is the Boltzmann constant, $T_0$ is the temperature (Kelvin), $P_{RIS}$ and $P_{BS}$ are the power of RIS and BS, respectively, $A_i=\frac{c^2}{16\cdot \pi^2 \cdot f^2 \cdot d^2_{(i,r^{,}v)}} (1-e^{-k(f) \cdot d_{(i,r^{,}v)}})$, where $d_{(i,r^{,}v)}$ is the distance between the network node $i$ and the related network node $r^{,}v$, $k(f)$ is the overall molecular absorption coefficient, and $I$ and $J$ are the set of RISs and of BSs, respectively, belonging to the multi-hop paths, transmitting the application traffic from the $A$ set:
\vspace{-0.44cm}
\begin{equation}\label{calI}
I = \{ r \in R \; | \; \exists \; r \in P, \forall \alpha \in A, s_{\alpha} \in \mathbb{S}_{\alpha}, P \in   \mathbb{P}_{s_{\alpha}} \},
\end{equation}
\vspace{-0.68cm}
\begin{equation}\label{calJ}
J = \{ b \in B \; | \; \exists \; b \in P, \forall \alpha \in A, s_{\alpha} \in \mathbb{S}_{\alpha}, P \in   \mathbb{P}_{s_{\alpha}} \}.
\end{equation}
Also in Eq. \eqref{capacitychanRIS}, the total channel gain between the RIS $r$ and the node $r^{,}v$ is given by \cite{8763780,9149411}: 
\begin{equation}\label{channelgain}
H_{(r,r^{,}v)} = \left ( \frac{c}{4 \cdot \pi \cdot d_{(r,r^{,}v)} \cdot f} \right)^2 \text{exp}(-d_{(r,r^{,}v)} \cdot k(f)).
\end{equation} $\phi_{(r,n_e,r^{,}v)}$ is the phase shift of reflecting element $n_e$ of RIS $r$ with respect to the related network node $r^{,}v$. $\psi_{(r,n_e,r^{,}v)}$ is the phase shift of the channel between the reflecting element $n_e$ of RIS $r$ and the node $r^{,}v$. For the optimal phase to maximize the transmission speed of each RIS \cite{8796365,9149411},  $\phi_{(r,n_e,r^{,}v)} = \psi_{(r,n_e,r^{,}v)}$. Finally, $Z$ is the set of reflecting elements of RIS $r$ used for reflecting signals from the RIS $r$ to the related node $r^{,}v$ over one link $(r,r^{,}v)$. The $Z$ set of reflecting elements are based on incoming links $e \in E_{in}(r)$ of the RIS $r$ that exists the application traffic $\alpha \in A$ with the data routed from the incoming links $e$ to the outgoing link $(r,r^{,}v)$, i.e., 
\begin{equation}\label{calY}
\begin{split}
Y= \{ e \in E_{in}(r) \; | &\; \exists \; e(r,r^{,}v) \in P, \\ &\forall \; \alpha \in A, s_{\alpha} \in \mathbb{S}_{\alpha}, P \in   \mathbb{P}_{s_{\alpha}} \},
\end{split}
\end{equation}
i.e., each incoming link $e \in E_{in}(r)$ of RIS $r$ satisfying Eq. \eqref{calY} will occupy one reflecting element $n_e$ for reflecting the signal to the related node $r^{,}v$. We assume that any $|Z|$ reflecting elements of RIS $r$ can be used to calculate the $C_{(r,r^{,}v)}(Z)$ from Eq. \eqref{capacitychanRIS}. Hence, the $Z$ set can be given by:
\begin{subnumcases}{Z=}
    N &, if $|Y| \geq |N|$ \label{calZ1}\\
\gamma &, if $|Y| < |N|$. \label{calZ2} 
\end{subnumcases}
Eq. \eqref{calZ1} means that with each RIS with $|N|$ reflecting elements, the maximum number of reflecting elements occupied is always $|N|$. In Eq. \eqref{calZ2}, the $\gamma$ set is given by:
\begin{equation} \label{calgamma}
\gamma = \{n_e \in N, \forall \; e \in Y \},
\end{equation}
where each incoming link $e \in Y$ (from Eq. \eqref{calY}) occupies one reflecting element $n_e$ from the $N$ set of reflecting elements.

We note that $E_{in}(r)$ and $E_{out}(r)$, for any RIS $r$, are defined by the network topology and that the set of possible paths $\mathbb{P}_{s_{\alpha}}$ is defined offline by the network management system. Thus, the parameters $X$ and $Y$ used in this model are upper-bounds and not necessarily the actual values for each network configuration. The higher the network load, the closer $X$ and $Y$ values will be to their actual values. Similarly, we have the reference parameters $I$ and $J$ defined offline by the network management system. Nevertheless, this modeling guarantees that each RIS will never be overloaded.


\begin{figure}
\centering
\includegraphics[width=7cm,height=4.6 cm]{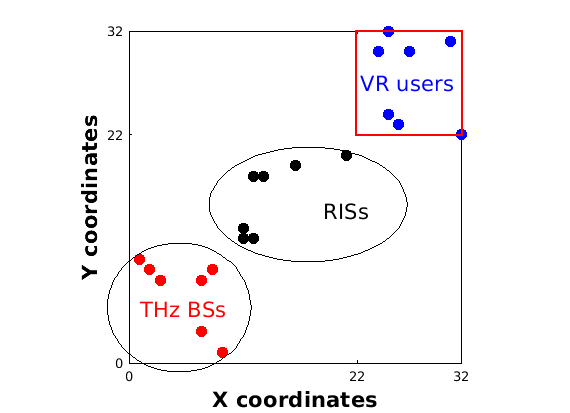}
\caption{RIS-assisted THz network in a flat area of $32 \times 32$ m, where VR users can move in the square area (red line).}\label{topology}
\vspace{-0.5cm}
\end{figure}

  
\begin{figure*}[ht]
  \centering
  \subfloat[Maximum multiplier $\lambda$.]{\includegraphics[ width=4.5cm, height=4.45cm]{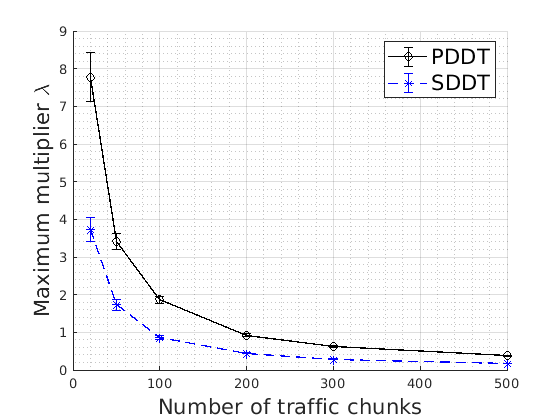}
  \label{lamda_static777}}
  \subfloat[Throughput gain.]{\includegraphics[ width=4.5cm, height=4.45cm]{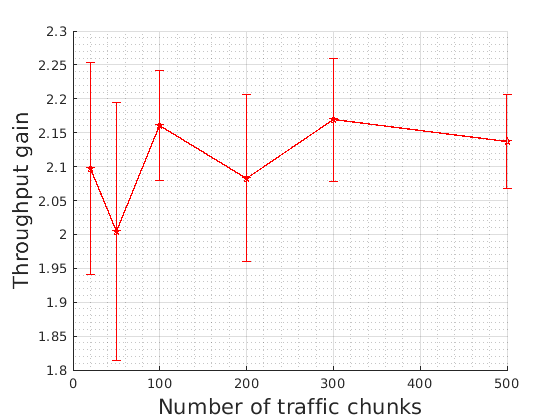}
  \label{throughputgain_static777}}
  \subfloat[Maximum multiplier $\lambda$.]{\includegraphics[ width=4.5cm, height=4.45cm]{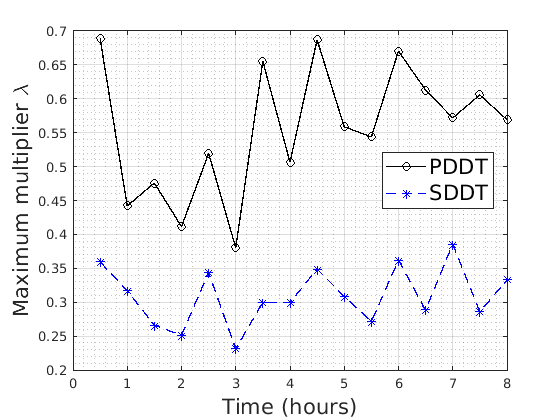}
  \label{lamda_dynamic777}}
  \subfloat[Throughput gain.]{\includegraphics[ width=4.5cm, height=4.45cm]{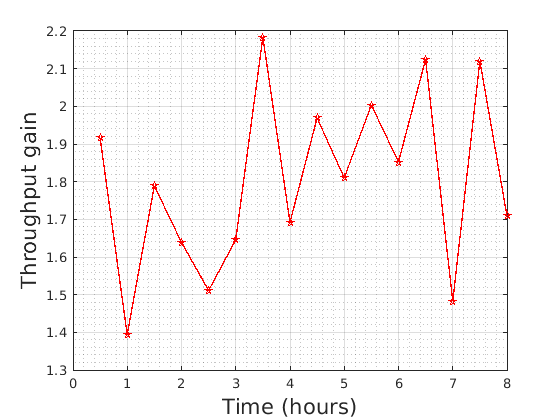}
  \label{throughputgain_dynamic777}}
  \caption{Results for maximum multiplier $\lambda$, and throughput gain in the RIS-assisted THz network, where (a),(b): VR users are static as presented in Fig. \ref{topology}; (c),(d): VR users randomly move every $30$ minutes in the square area (red line) of Fig. \ref{topology}.}
  \label{results}
  \vspace{-0.67cm}
  \end{figure*}

\section{Numerical results}\label{Numerre}
In this section, we show the evaluation results in the
smart factory downlink application scenario, as illustrated in Fig. \ref{arch} and the PDDT. Assume that ARCM always has the current RIS-assisted THz network topology and updates the application traffic set $A$ from THz BSs, where each application traffic $\alpha \in A$ is split into the fixed traffic chunks of $t_{s_{\alpha}}=0.5$ Gbit, and these chunks are randomly assigned to any BSs. 
  
Assume that ARCM has the current RIS-assisted THz topology in a flat area of $32 \times 32$ m, shown in Fig. \ref{topology}, whereby the coordinates of THz BSs, RISs, and VR users are: $1 \leq x,y \leq 10$ m, $11 \leq x,y \leq 21$ m, and $22 \leq x,y \leq 32$ m, respectively. Assume that two nodes in the range of $20$ m can be connected together, and the RIS controller only allows transmitting the data up to this distance, otherwise we need the RISs to forward the traffic from the THz BSs. The THz BSs and RISs are located as static, while VR users can move. The values used in the evaluation are summarized in Table \ref{tab:table1}. 


Now, we analyze the results of maximum multiplier $\lambda$ over the different $D$ sets of $t_{s_{\alpha}}$ traffic chunks in Fig. \ref{lamda_static777}, which allows us to compare the methods of parallel downlinks (PDDT) and single downlinks (SDDT). $|\mathbb{P}_{s_{\alpha}}|=5$ possible shortest paths for  PDDT, and $|\mathbb{P}_{s_{\alpha}}|=1$ shortest path for the SDDT are chosen from the topology in Fig. \ref{topology}. Assume that VR users are standing at the certain positions as in Fig. \ref{topology} under the consideration of $|D|=20, 50, 100, 200, 300, 500$ $t_{s_{\alpha}}$ traffic chunks. We use Table \ref{tab:table1}, and the number of chunks from the $x$-axis of Fig. \ref{lamda_static777} as the inputs to apply for the LP in subsection \ref{LP} and get the maximum multiplier $\lambda$ values on the $y$-axis as the outputs, whereby the $s_{\alpha}$ source nodes assigned to $t_{s_{\alpha}}$ chunks  and the $d_{\alpha}$ destination nodes are randomly chosen. We run $10$ random independent replications to obtain the average values with confidence intervals of $95\%$, shown in Fig. \ref{lamda_static777}.

 \begin{table}[t!]
  \centering
    \vspace{0.4 cm}
  \caption{List of main values applied for evaluation.}
  \label{tab:table1}
  \begin{tabular}{ll}
    \toprule
   Values & Meaning\\
    \midrule
    $|B|=7$ & Number of THz BSs.\\
    $|R|=7$ & Number of RISs.\\
    $|V|=7$ & Number of VR users.\\
    $|N| =16$ & Number of reflecting elements for each RIS.\\
    $t_{s_{\alpha}}=0.5$ & Traffic chunk assigned for source node $s_{\alpha}$ in Gbit.\\
    $P_{BS}=10$ & Power of each BS in Watt.\\
    $P_{RIS}=1$ & Power of each RIS in Watt.\\
    $T_o=300$ & Temperature in Kelvin.\\
    $\tau = 0.5$ & Time observed for the queuing in seconds.\\
    $L=10$ & Queuing length of each RIS in Gbits.\\
    $|\mathbb{P}_{s_{\alpha}}|=5$ & Parallel  downlink data distribution with $5$ paths.\\
     $|\mathbb{P}_{s_{\alpha}}|=1$ & Serial downlink data distribution.\\
    $f=1$ & Operational center frequencies around $1$ THz.\\
    $k(f)= 0.0016$  & Molecular absorption coefficient in $m^-1$.\\
    $W=3$ & Bandwidth in GHz.\\
     \bottomrule
  \end{tabular}\vspace{-0.76cm}
\end{table}

As expected, Fig. \ref{lamda_static777} confirms that the PDDT achieves a greater end-to-end throughput (maximum multiplier $\lambda$) compared to SDDT. The reason is because the PDDT finds the paths that have fewer overloaded RISs.  PDDT outperforms  SDDT for all $D$ sets of $t_{s_{\alpha}}$ traffic chunks. The throughput decreases, when the number of $t_{s_{\alpha}}$ traffic chunks increases because it becomes more difficult to find paths without overloaded RISs. Note that if $\lambda \geq 1$, each traffic chunk $t_{s_{\alpha}}$ can be adequately transmitted in the network. If $\lambda > 1$, the network can still support more traffic; in other words, we are using only part of the available resources. Otherwise, for $\lambda < 1$, the network cannot transfer the demands, i.e., $t_{s_{\alpha}}$ needs to be less than $0.5$ Gbit. In this situation, the control plane can inform the THz system that the traffic chunk demands cannot be satisfied. Then, it is necessary to decide if the demands will be blocked or served with less Quality of Service (QoS). To better understand this discussion, we define the maximum traffic chunk as $t_{s_{\alpha}}^{max}= t_{s_{\alpha}} \cdot \lambda$.
For example, at $|D|=100$ traffic chunks for the PDDT, we have the average value $\lambda = 1.86089 > 1$ in Fig. \ref{lamda_static777}, then the corresponding maximum traffic chunk is equal to $t_{s_{\alpha}}^{max}=0.930445  > 0.5$ Gbit. At $|D|=300$ traffic chunks for the PDDT, we have the average value $\lambda = 0.62776 < 1$ in Fig. \ref{lamda_static777}, then $t_{s_{\alpha}}^{max}=0.31388 < 0.5$ Gbit. At $|D|=200$  chunks for the PDDT, we have the average value $\lambda \approx 1$ in Fig. \ref{lamda_static777}, then $t_{s_{\alpha}}^{max} \approx 0.5$ Gbit. Hence, we can conclude $|D|=200$ chunks for the PDDT is a good case for transmitting the traffic chunks $t_{s_{\alpha}}=0.5$ Gbit because this case can approximately use the full network resource without wasting resource. For SDDT, the traffic chunks with $t_{s_{\alpha}} = 0.5$ Gbit can only be smoothly transmitted in the network, i.e., $\lambda \geq 1$, if $|D| < 100$ chunks. Next, we show the throughput gain $G$ in Fig. \ref{throughputgain_static777}, defined as the ratio of the maximum multiplier obtained by the PDDT to the maximum multiplier obtained by the SDDT: $G = \frac{\lambda_{PDDT}}{\lambda_{SDDT}}$.
By using the values in Fig. \ref{lamda_static777}, we get throughput gain $G$ in Fig. \eqref{throughputgain_static777}.  On average, we get the significant throughput gain if using the PDDT. 

Next, we consider the dynamic scenario in Fig. \ref{lamda_dynamic777} and Fig. \ref{throughputgain_dynamic777}. All system configurations are similar to Fig. \ref{results}, but we consider VR user moves in the coordinates: $22 \leq x,y \leq 32$ m. The network connectivity and topology changes every $30$ minutes. We consider $|D|=300$ $t_{s_{\alpha}}$ chunks, whereby the $s_{\alpha}$ sources assigned to $t_{s_{\alpha}}$ chunks  and the $d_{\alpha}$ destinations are randomly chosen every $30$ minutes. In general, we observe that the throughput of PDDT still outperforms the SDDT. There are result fluctuations over time due to the position changes of VR users and the changes of transmitting and receiving nodes. We get a throughput gain of $1.4$ at $1$ hour in the worst case and up to nearly $2.2$ at $3.5$ hours in the best case if using the PDDT. 

In summary, we observe that if $\lambda \geq 1$, the size of each traffic chunk $t_{s_{\alpha}}$ in the system can be  transmitted with sufficient performance. If the network cannot serve the traffic requests as demanded, i.e., $\lambda < 1$, it is necessary to decide if the demand will be blocked or served with less QoS. In the case of when $\lambda \geq 1$, we conclude that using the SDDT can still be used when the overall traffic in the system is low, e.g., $|D|<100$  traffic chunks (Fig. \ref{lamda_static777}). SDDT is less complex and there are no issues of differential delay typical for multipathing, as it practically would be the case with PDDT. However, as the traffic increases, e.g., $|D|\geq 100$ chunks (Fig. \ref{lamda_static777}), PDDT is a better choice due to its flexibility in distributing data over parallel downlinks which then always outperforms SDDT.

\vspace{-0.11cm}
\section{Conclusion}\label{concl}
We proposed a parallel  downlink data distribution system and developed multi-criteria optimization solutions that can improve throughput, while balancing allocation of RIS devices in indoor THz networks. The results showed that the proposed system can enhance the performance in terms of throughput, as compared to serial download link distribution, especially under (controlled) mobility conditions of users. Parallel downlink distribution is able to balance load among RISs in the networks, which was our main goal.  Our analysis showed that a tradeoff can be found between the RIS utilization and traffic distribution in parallel chunks. Future work requires reconsideration of some of the strong assumptions, such as the phase shifts of RISs and interference. An interesting case can be the management of more general mobility in an indoor setting where PDDT is expected to yield superior performance.
\section*{Acknowledgment}
We acknowledge partial support of the Federal Ministry of Education and Research, Germany, project 6G-RIC, 16KISK031, and  DFG Project Nr. JU2757/12-1.
\bibliographystyle{IEEEtran}

\bibliography{nc-rest}

\end{document}